\documentclass[12pt]{article}
\pdfoutput=1
\usepackage{subfigure}
\usepackage{amssymb,amsmath}
\usepackage{graphicx}
\usepackage{color}
\usepackage[colorlinks=true
,urlcolor=blue
,citecolor=blue
,linkcolor=blue
,pagecolor=blue
,linktocpage=true
,pdfproducer=medialab
]{hyperref}
\usepackage[a4paper]{geometry}
\usepackage{placeins}

\makeatletter \renewcommand{\@dotsep}{10000} \makeatother

\def\tst{\tilde t}
\def\ttau{\tilde \tau}

\mathchardef\mhyphen="2D

\newcommand{\beq}{\begin{equation}}
\newcommand{\eeq}{\end{equation}}
\newcommand{\bea}{\begin{eqnarray}}
\newcommand{\eea}{\end{eqnarray}}

\newcommand\prd[3]{{\it Phys.\ Rev.\ }{\bf D #1} (#2) #3}
\newcommand\prl[3]{{\it Phys.\ Rev.\ Lett.\ }{\bf #1} (#2) #3}

\newcommand\jhep[3]{{\it J. High Energy Phys.\ }{\bf #1} (#2) #3}

\usepackage[nodisplayskipstretch]{setspace}

\begin{document}

\begin{titlepage}
\pagestyle{empty}

\vspace*{0.2in}
\begin{center}
{\bf Light Higgsinos, Heavy Gluino and $b-\tau$ Quasi-Yukawa Unification: \\ Will the LHC find the Gluino?
  }\\
\vspace{1cm}
{\bf  Aditya Hebbar$^{a,}$\footnote{E-mail: adityah@udel.edu}, Qaisar Shafi $^{a,}$\footnote{E-mail: shafi@bartol.udel.edu} and
Cem Salih $\ddot{\rm U}$n$^{b,}\hspace{0.05cm}$\footnote{E-mail: cemsalihun@uludag.edu.tr}}
\vspace{0.5cm}

{\it
$^a$Bartol Research Institute, Department of Physics and Astronomy,\\
University of Delaware, Newark, DE 19716, USA \\
$^b$Department of Physics, Uluda\~{g} University, TR16059 Bursa, Turkey
}

\end{center}

\vspace{0.5cm}
\begin{abstract}
\noindent
A wide variety of unified models predict asymptotic relations at $M_{GUT}$ between the b quark and $\tau$ lepton Yukawa couplings. Within the framework of supersymmetric SU(4) $\times$ SU(2)$_L \times$ SU(2)$_R$, we explore regions of the parameter space that are compatible with b-$\tau$ quasi-Yukawa unification and the higgsinos being the lightest supersymmetric particles ($\lesssim$ 1 TeV). Among the colored sparticles, the stop weighs more than 1.5 TeV or so, whereas the squarks of the first two families are signifcantly heavier, approaching 10 TeV in some cases. The gluino mass is estimated to lie in the 2-4 TeV range which raises the important question: Will the LHC find the gluino? 

\end{abstract}

\end{titlepage}

%\baselineskip 36pt

%%%%%%%%%%%%%%%%%%%%%%%%%%s
% Main body
%%%%%%%%%%%%%%%%%%%%%%%%%%

\section{Introduction}
\label{ch:introduction}

Low scale supersymmetry (SUSY) remains an attractive extension of the Standard Model despite the apparent absence thus far of any direct experimental signature for its existence, or any new physics for that matter, at the LHC \cite{LHC}. A supersymmetric scenario consisting of relatively light ( $\lesssim$ 1 TeV higgsinos) has attracted some attention in recent years \cite{Drees:1996pk, Baer:2012mv} . It has been emphasized that this domain of supersymmetry may be more readily accessible at the much discussed ILC rather than at the LHC. The parameter space in this case has been referred to as `natural' SUSY \cite{Baer:2012mv}, where the Minimally Supersymmetric Standard Model (MSSM) $\mu$ parameter and related parameters associated with radiative electroweak symmetry breaking (REWSB) are restricted to be comparable in magnitude to the Z-boson mass.

Our motivation in this paper is to realize a supersymmetric scenario with light higgsinos within the framework of unified models that also displays approximate third family Yukawa coupling unification (YU). Well-known examples include approximate t-b-$\tau$ Yukawa unification \cite{big-422, bigger-422} in SO(10)\cite{SO(10)} and SU(4) $\times$ SU(2)$_L$ $\times$ SU(2)$_R$ (4-2-2)\cite{pati-salam}, and b-$\tau$ Yukawa unification which can occur in SO(10), 4-2-2 and SU(5) models. t-b-$\tau$ YU requires that the two MSSM Higgs doublets (H$_d$ and H$_u$) reside in the 10 - plet of SO(10). However, to incorporate fermion masses and mixings, it is necessary to extend the Higgs sector by introducing additional Higgs fields in the (15,1,3) and/or the (15,1,1) representations of 4-2-2. This breaks exact Yukawa unification, but the deviation from Yukawa unification can be restricted to within 20\%, and such a modified scheme is often referred to as Quasi-Yukawa Unification (QYU)\cite{Gomez:2002tj, Dar:2011sj}. If Higgs fields from the above two (and possibly other) representations are present, then the top quark Yukawa coupling, in particular, may receive large corrections and therefore no longer unify with the b-$\tau$ Yukawa couplings. This particular scenario, called b-$\tau$ QYU, will be the focus of our study in this work. Note that TeV scale supersymmetry plays a critical role via radiative corrections \cite{bartol2} in implementing approximate Yukawa unification at M$_{GUT}$. This may be considered additional evidence in support of supersymmetric Grand Unified Theories (GUTs) versus their non-supersymmetric counterparts which do not possess such threshold corrections.

The supersymmetric 4-2-2 with left-right symmetry naturally allows for non-universality in the MSSM gaugino sector. Thus, we can write
\begin{equation}
M_{1}=\frac{3}{5}M_{2}+\frac{2}{5}M_{3}
\label{gauginomasses}
\end{equation}
which is implied by LR symmetry and hypercharge composition:
\begin{equation}
M_{R}=M_{L}\equiv M_{2}~,\hspace{0.5cm} Y=\sqrt{\frac{3}{5}}I_{3R}+\sqrt{\frac{2}{5}}(B-L),
\label{LRsymm}
\end{equation}
where $M_{1}$, $M_{2}$ and $M_{3}$ are the asymptotic soft supersymmetric breaking (SSB) gaugino masses for $U(1)_{Y}$, $SU(2)_{L}$ and $SU(3)_{C}$. For the scalar sector we work with the so-called Non-Universal Higgs Model 2 (NUMH 2) structure in which the soft scalar masses associated with the sfermions and the two MSSM Higgs doublets are treated as independent parameters.

In order to explore the parameter space compatible with quasi b-$\tau$ QYU and light higgsino, we employ the the fine-tuning parameter $\Delta_{EW}$ defined in \cite{Baer:2012mv}

\begin{equation} \Delta_{EW} \equiv \rm{max}_i \left(C_i\right)/(M_Z^2/2) ,
\end{equation}
where $C_{H_u}=|-m_{H_u}^2\tan^2\beta /(\tan^2\beta -1)|$,
$C_{H_d}=|m_{H_d}^2/(\tan^2\beta -1)|$ and $C_\mu =|-\mu^2|$, along
with analogous definitions for $C_{\Sigma_u^u(k)}$ and
$C_{\Sigma_d^d(k)}$. We further constrain the parameter space by requiring that $\Delta_{EW}$ $\lesssim$ 200. Note that this condition is compatible with the light higgsino condition $\mu \lesssim 1 $ TeV.

If the neutralino is required to be the lightest supersymmetric particle (LSP), then the range of $\Delta_{EW}$ we have considered here will result in mostly higgsino-like LSP. For smaller values of $\Delta_{EW}$, say of order 25-50, a second dark matter component, such as an axion, is needed to satisfy the dark matter abundance reported by WMAP \cite{WMAP9}. Values close to the upper limit (where $\mu \sim$ 1 TeV) yield solutions in which the higgsino alone satisfies the observed dark matter relic abundance in the universe. 

In our discussion, following previous work \cite{Dar:2011sj}, we express QYU as follows:
\begin{equation}
y_{t}:y_{b}:y_{\tau}=|1+C_{t}|:|1-C_{b\tau}|:|1+3C_{b\tau}|,
\label{QYUcon}
\end{equation}
where $C_{t}$ measures deviation in $y_{t}$, while $C_{b\tau}$ measures the deviation of $y_{b}$ and $y_{\tau}$. The factor of 3 in Eq.(\ref{QYUcon}) has its origin in the Clebsch-Gordon coefficient associated with the 15-dimensional SU(4)$_C$ representation \cite{Gomez:2002tj,Lazarides:1980nt}. Note that $C_{t}$ does not have to be related to $C_{b\tau}$. For definiteness, we restrict $C_{b\tau} \leq$ 0.2 and $C_t$ $\leq$ 2.    

The rest of the paper is organized as follows: In Section \ref{sec:scan} we describe our scanning procedure and summarize the experimental constraints employed in our analysis. In section \ref{sec:fund} we discuss the regions in the fundamental parameter space which are compatible with QYU and the light higgsino conditions. In Section \ref{sec:spec} we present  the sparticle mass spectrum and its implications for fine-tuning and DM. Section \ref{sec:DM} focuses on the implications for dark matter (DM) based on current results from  direct detection experiments. Our conclusions are summarized in \ref{sec:conc}.

\section{Scanning Procedure and Experimental Constraints}
\label{sec:scan}

We employ the ISAJET~7.84 package~\cite{ISAJET} 
 to perform random scans over the parameter space 
 given below. 
In this package, the weak scale values of gauge and third 
 generation Yukawa couplings are evolved to 
 $M_{\rm GUT}$ via the MSSM renormalization group equations (RGEs)
 in the $\overline{DR}$ regularization scheme.
We do not strictly enforce the unification condition
 $g_3=g_1=g_2$ at $M_{\rm GUT}$, since a few percent deviation
 from unification can be assigned to unknown GUT-scale threshold
 corrections~\cite{Hisano:1992jj}.
With the boundary conditions given at $M_{\rm GUT}$, 
 all the SSB parameters, along with the gauge and Yukawa couplings, 
 are evolved back to the weak scale $M_{\rm Z}$.

In evaluating Yukawa couplings, the SUSY threshold 
 corrections~\cite{Pierce:1996zz} are taken into account 
 at the common scale $M_{\rm SUSY}= \sqrt{m_{\tst_L}m_{\tst_R}}$. 
The entire parameter set is iteratively run between 
 $M_{\rm Z}$ and $M_{\rm GUT}$ using the full 2-loop RGEs
 until a stable solution is obtained.
To better account for leading-log corrections, one-loop step-beta
 functions are adopted for gauge and Yukawa couplings, and
 the SSB parameters $m_i$ are extracted from RGEs at appropriate scales
 $m_i=m_i(m_i)$.
The RGE-improved 1-loop effective potential is minimized
 at an optimized scale $M_{\rm SUSY}$, which effectively
 accounts for the leading 2-loop corrections.
Full 1-loop radiative corrections are incorporated
 for all sparticle masses.

The requirement of radiative electroweak symmetry breaking
 (REWSB)~\cite{Ibanez:1982fr} puts an important theoretical
 constraint on the parameter space.
Another important constraint comes from limits on the cosmological
 abundance of stable charged particles~\cite{Olive}.
This excludes regions in the parameter space where charged
 SUSY particles, such as $\ttau_1$ or $\tst_1$,
 become the LSP.

We have performed random scans 
 for the following parameter space:

\begin{eqnarray}
\label{parameterRange}
0 \leq  m_{0}  \leq 20~ {\rm TeV} \nonumber \\
0 \leq  M_{2}  \leq 5~ {\rm TeV} \nonumber \\ 
0 \leq  M_{3}  \leq 5~ {\rm TeV} \nonumber \\
-3 \leq A_{0}/m_{0} \leq 3 \\
2 \leq \tan\beta \leq 60 \nonumber \\
 0 \leq  m_{H_{u}}  \leq 20~ {\rm TeV} \nonumber \\
  0 \leq  m_{H_{d}}  \leq 20~ {\rm TeV} , \nonumber
  \label{paramSpace}
\end{eqnarray}

\noindent with  $\mu > 0$ and  $m_t = 173.3\, {\rm GeV}$  \cite{:2009ec}.
Note that our results are not too sensitive to one 
 or two sigma variation in the value of $m_t$  \cite{bartol2}.
We use $m_b^{\overline{DR}}(M_{\rm Z})=2.83$ GeV 
 which is hard-coded into ISAJET.

In scanning the parameter space, we employ the Metropolis-Hastings
 algorithm as described in \cite{Belanger:2009ti}. 
The data points collected all satisfy the requirement of REWSB, 
 with the neutralino being the LSP in each case. 
After collecting the data, we impose the mass bounds on 
 all the particles \cite{Olive} and 
 use the IsaTools package
 to implement the following phenomenological constraints ~\cite{bsg, bmm,mamoudi}: 

\begin{eqnarray} 
m_h  = 123-127~{\rm GeV}~~& 
\\
m_{\tilde{g}}\geq 1.8~{\rm TeV}\\
0.8\times 10^{-9} \leq{\rm BR}(B_s \rightarrow \mu^+ \mu^-) 
  \leq 6.2 \times10^{-9} \;(2\sigma)~~&& 
\\ 
2.99 \times 10^{-4} \leq 
  {\rm BR}(b \rightarrow s \gamma) 
  \leq 3.87 \times 10^{-4} \; (2\sigma)~~&&  
\\
0.15 \leq \frac{
 {\rm BR}(B_u\rightarrow\tau \nu_{\tau})_{\rm MSSM}}
 {{\rm BR}(B_u\rightarrow \tau \nu_{\tau})_{\rm SM}}
        \leq 2.41 \; (3\sigma)~~&&  
\\
 0.0913 \leq \Omega_{\rm CDM}h^2 (\rm WMAP9) \leq 0.1363   \; (5\sigma)~~&\cite{WMAP9}&.
%\\ 
% 2.1 \times 10^{-10} \leq \Delta a_{\mu} 
%  \leq 50.1 \times 10^{-10} \; (3\sigma)~~&\cite{BNL}&
%\labels{constraints}
\end{eqnarray} 
We emphasize the mass bounds on the Higgs boson \cite{ATLAS, CMS}, and the gluino \cite{Aad:2012tfa}, since the experiments at the Large Hadron Collider (LHC) have had a strong impact on the bounds on these particles. The rare $B-$meson decays have a strong impact on the parameter space, since the SM predictions are already in good agreement with the experimental results. We have applied the constraints from ${\rm BR}(B_s \rightarrow \mu^+ \mu^-) $ \cite{Aaij:2012nna} and ${\rm BR}(b \rightarrow s \gamma) $ \cite{Amhis:2012bh} within $2\sigma$ uncertainty, while the MSSM predictions on  ${\rm BR}(B_u\rightarrow \tau \nu_{\tau})$ are limited to within $3\sigma$ uncertainty \cite{Asner:2010qj}. 

Another strict constraint comes from the DM observables. Since the LSP is proposed as a candidate for DM, the regions in the fundamental parameter space which yield charged sparticles as LSP are excluded. Thus, we accept only those solutions for which one of the neutralinos is the lightest supersymmetric particle (LSP) without necessarily saturating the $5\sigma$ dark matter relic abundance bound observed by WMAP9. This is due to the fact that we are primarily motivated by `natural' SUSY which we interpret to mean MSSM $\mu$ parameter $\lesssim$ 1 TeV. The LSP higgsino in this case does not necessarily saturate the DM abundance. The impact of direct detection on the parameter space is discussed in Section \ref{sec:DM}. 

Finally, as far as the muon anomalous magnetic moment $a_{\mu}$ is concerned, we require that the solutions must be at least as consistent with the data as the Standard Model ($0 \leq \Delta a_{\mu} \leq 55.4\times 10^{-10}$ \cite{Bennett:2006fi}).

\section{Parameter Space compatible with Quasi-Yukawa Unification}
\label{sec:fund}
\begin{figure}[h!]
\centering
\subfigure{\includegraphics[scale=1]{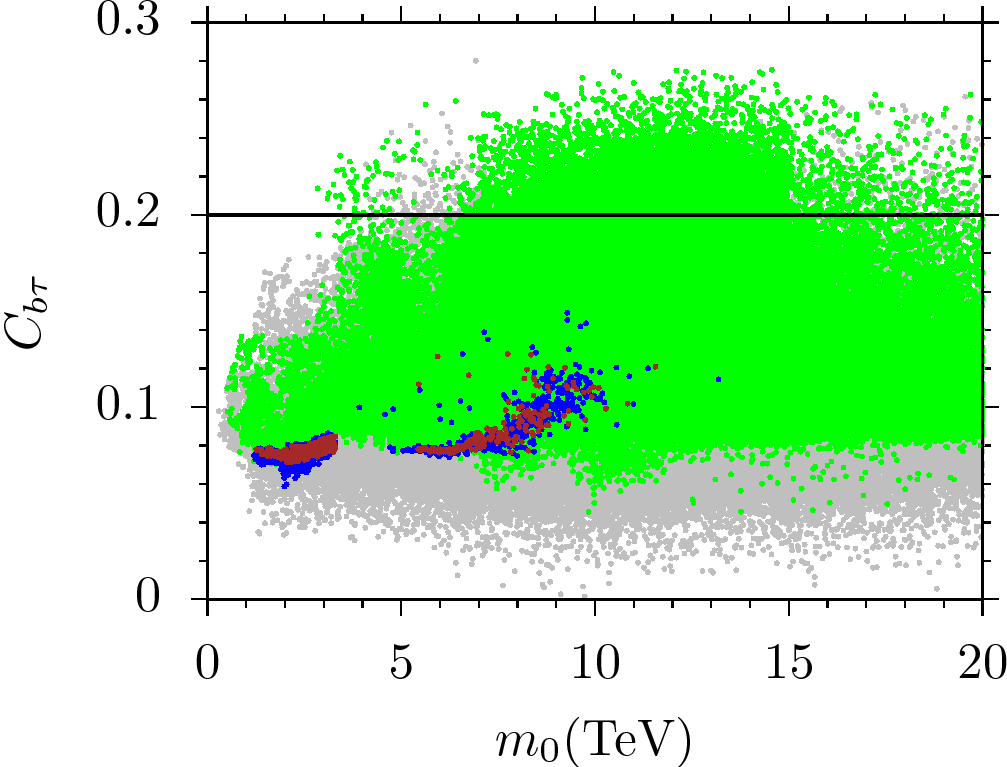}}
\subfigure{\includegraphics[scale=1]{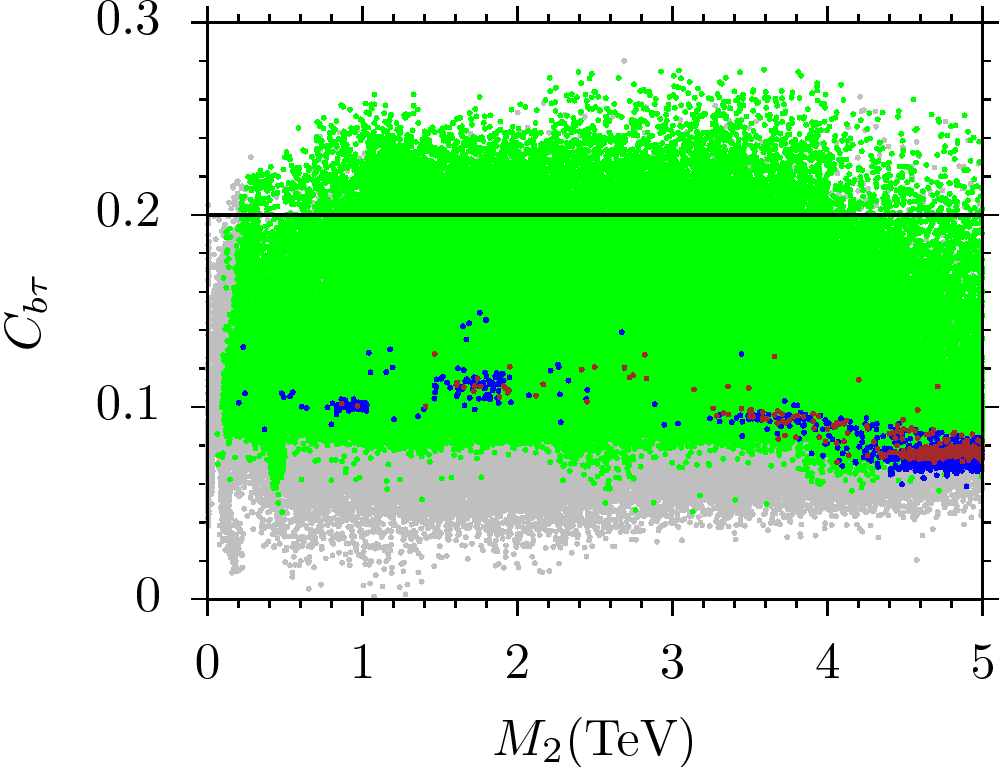}}
\subfigure{\includegraphics[scale=1]{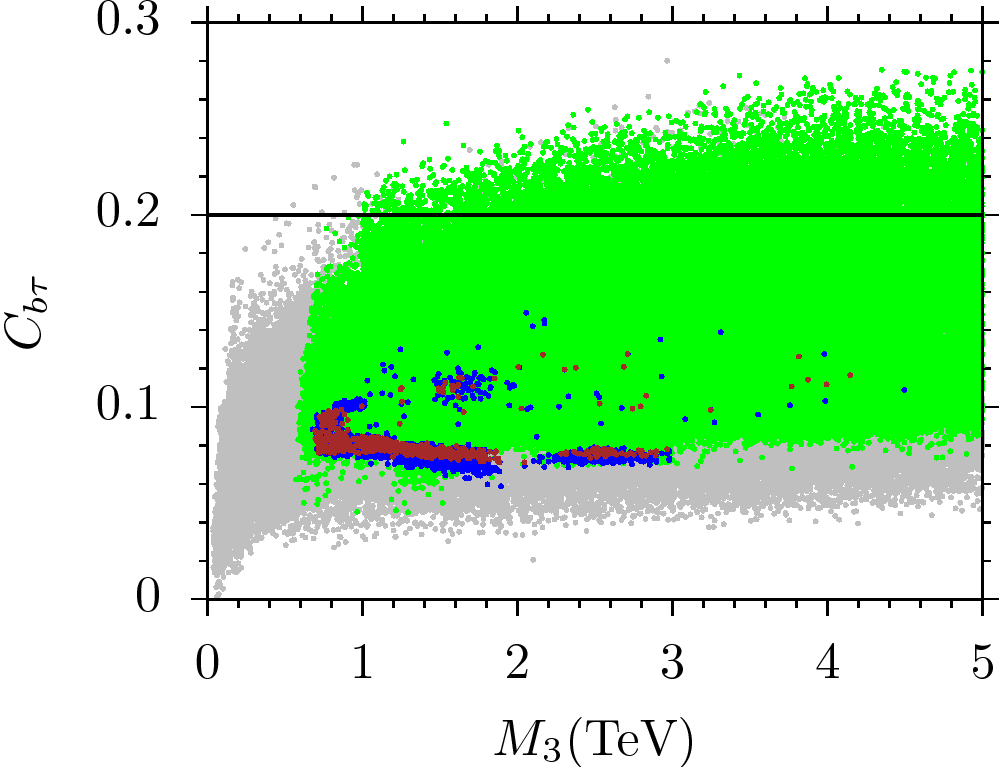}}
\subfigure{\includegraphics[scale=1]{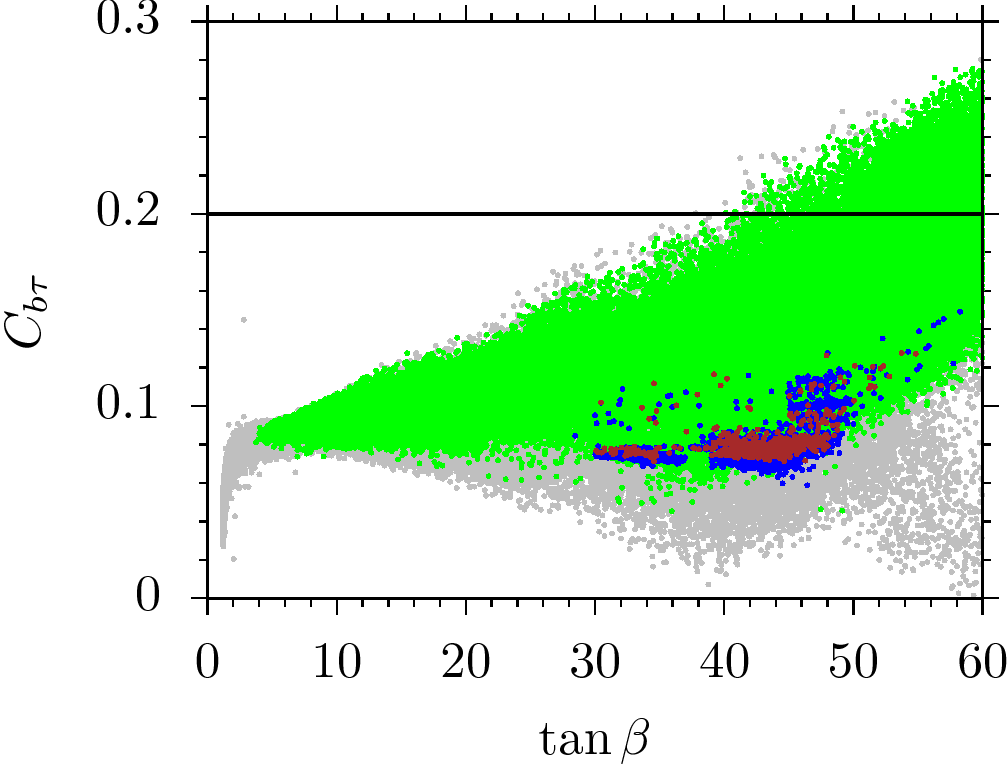}}
\caption{Plots in the $C_{b\tau}-m_{0}$, $C_{b\tau}-M_{2}$, $C_{b\tau}-M_{3}$ and $C_{b\tau}-\tan\beta$ planes. All points are compatible with REWSB and neutralino LSP. Green points satisfy the experimental constraints. Blue points form a subset which is compatible with $b-\tau$ QYU, $\mu \lesssim$ 1 TeV and $\Delta_{EW} < 200$. Brown points are a subset of blue and represent solutions consistent with the WMAP bound.}
\label{fig1}
\end{figure}

In this section we present the fundamental parameter space of $b-\tau$ QYU and discuss its impact on the low scale. Figure \ref{fig1} displays $C_{b\tau}$ vs. the fundamental parameters with plots in $C_{b\tau}-m_{0}$, $C_{b\tau}-M_{2}$, $C_{b\tau}-M_{3}$ and $C_{b\tau}-\tan\beta$ planes. All points are compatible with REWSB and neutralino LSP. Green points satisfy all experimental constraints. Blue points form a subset which is compatible with $\mu \lesssim$ 1 TeV and $\Delta_{EW} < 200$. Brown points are a subset of blue and represent solutions consistent with the WMAP bound on the relic abundance of the LSP neutralino. We do not apply $C_{b\tau} \leq 0.2$, but instead indicate this bound with a horizontal line. As seen from the $C_{b\tau}-m_{0}$, $C_{b\tau}-M_{2}$ and $C_{b\tau}-M_{3}$ plots, most of the solutions are below the horizontal line at $C_{b\tau}$ = 0.2 and hence, $b-\tau$ QYU is not a strong constraint on these parameters. The plots show that while $m_{0}$ cannot be lower than about 2 TeV or heavier than 10 TeV, $M_{2}$ and $M_{3}$ can be as low as about 800 GeV. However, we note that all the solutions with $C_{b\tau} < 0.1 $ have $M_2 > 3 $ TeV and $M_3 < 2$ TeV. The $C_{b\tau}-\tan\beta$ plot shows that the fine-tuning condition requires $\tan\beta \gtrsim 30$.

\begin{figure}[h!]
\centering
\subfigure{\includegraphics[scale=1]{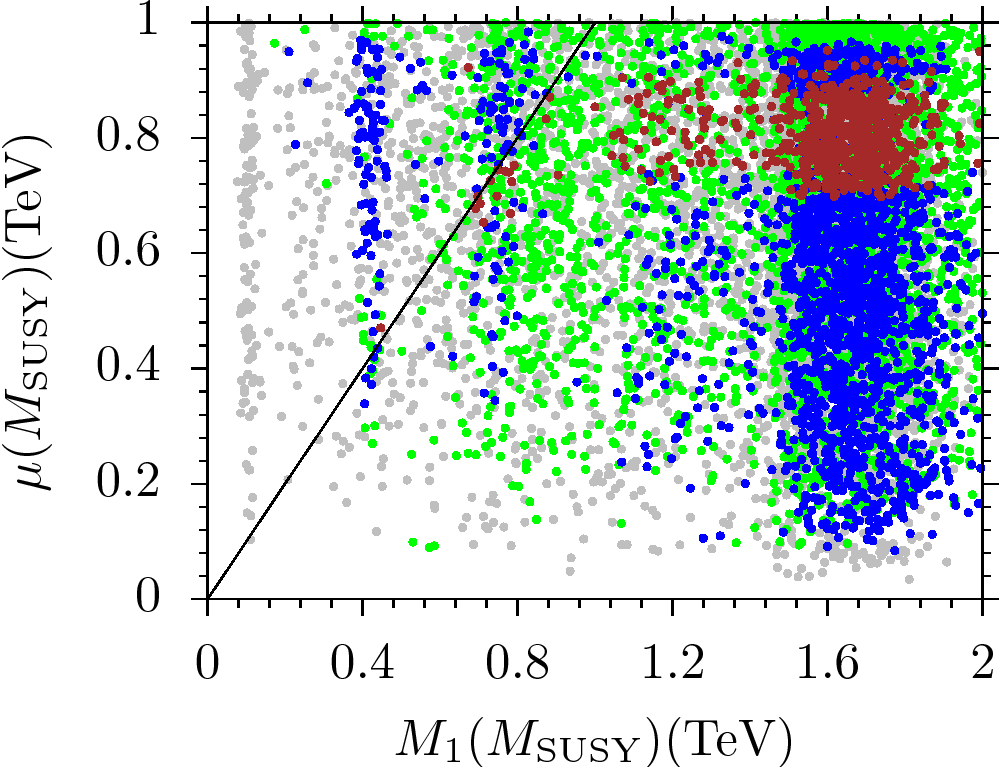}}
\subfigure{\includegraphics[scale=1]{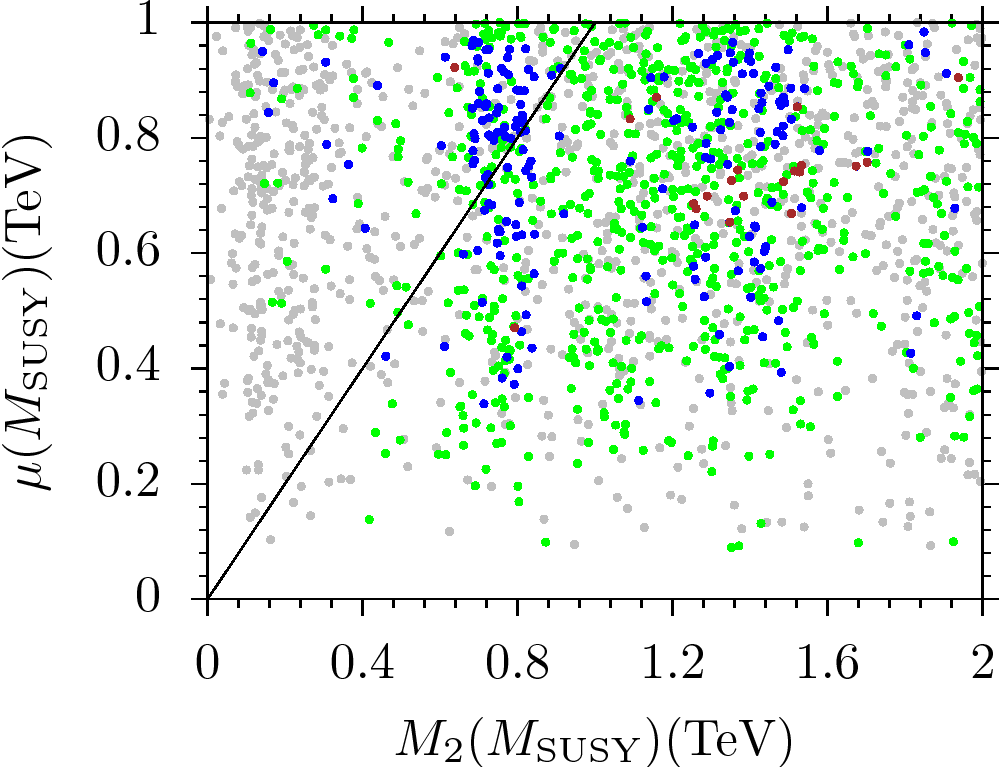}}
\caption{Plots in the $M_{1}-\mu$, $M_{2}-\mu$ planes, which represent the low scale values of these parameters. Color coding is the same as in Figure \ref{fig1} except that the blue points now satisfy $C_{b\tau} \leq 0.2$ in addition to $\mu \lesssim 1$ TeV and $\Delta_{EW} < 200$. The lines indicate the regions where $M_{1}=\mu$ (left) and $M_{2}=\mu$ (right) }
\label{fig2}
\end{figure}

We continue discussing the fundamental parameters in Figure \ref{fig2} with plots in $M_{1}-\mu$, $M_{2}-\mu$ planes, which represent the low scale values of these parameters. The color coding is the same as Figure \ref{fig1}, except that the blue points now satisfy $C_{b\tau} < 0.2$ in addition to $\mu \lesssim 1 $ TeV and $\Delta_{EW} < 200$. These parameters simply show the masses of neutralinos at the low scale, since $M_{1}$ and $M_{2}$ are the SSB mass terms for bino and wino respectively, while $\mu$ determines masses of the higgsinos.  Except for a few points near the line in the two plots which indicate a higgsino-bino or higgsino-wino mixture dark matter, we see that for much of the parameter space, the dark matter is composed mainly of higgsinos. 

\begin{figure}[h!]
\centering
\subfigure{\includegraphics[scale=0.9]{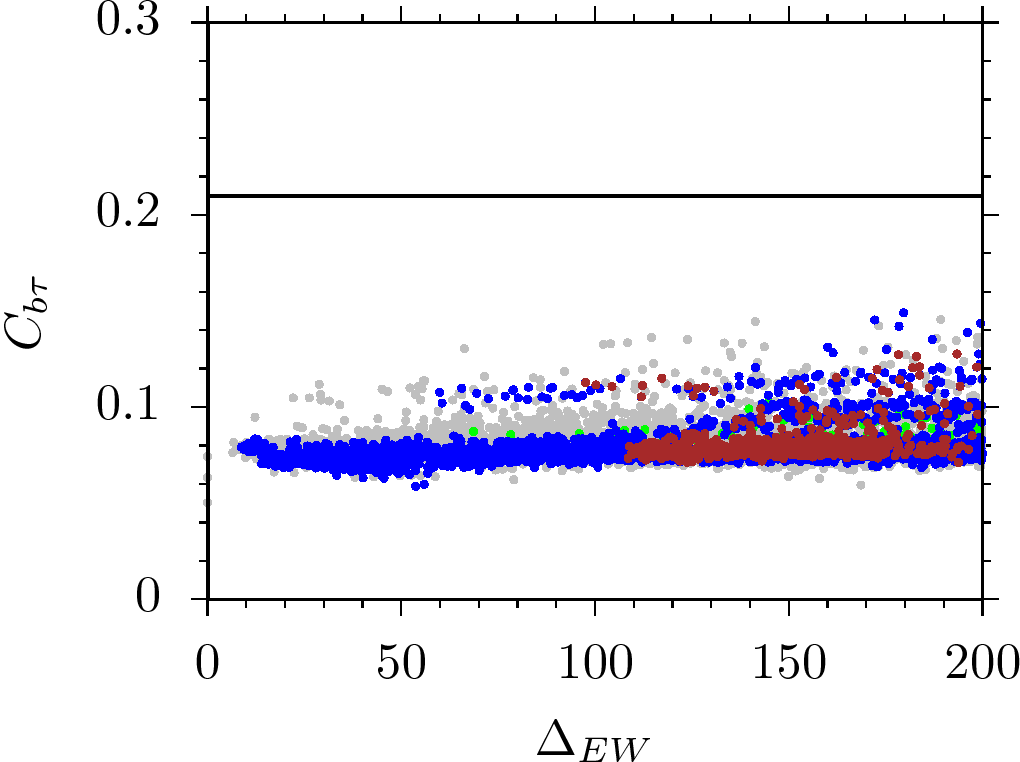}}
\subfigure{\includegraphics[scale=0.9]{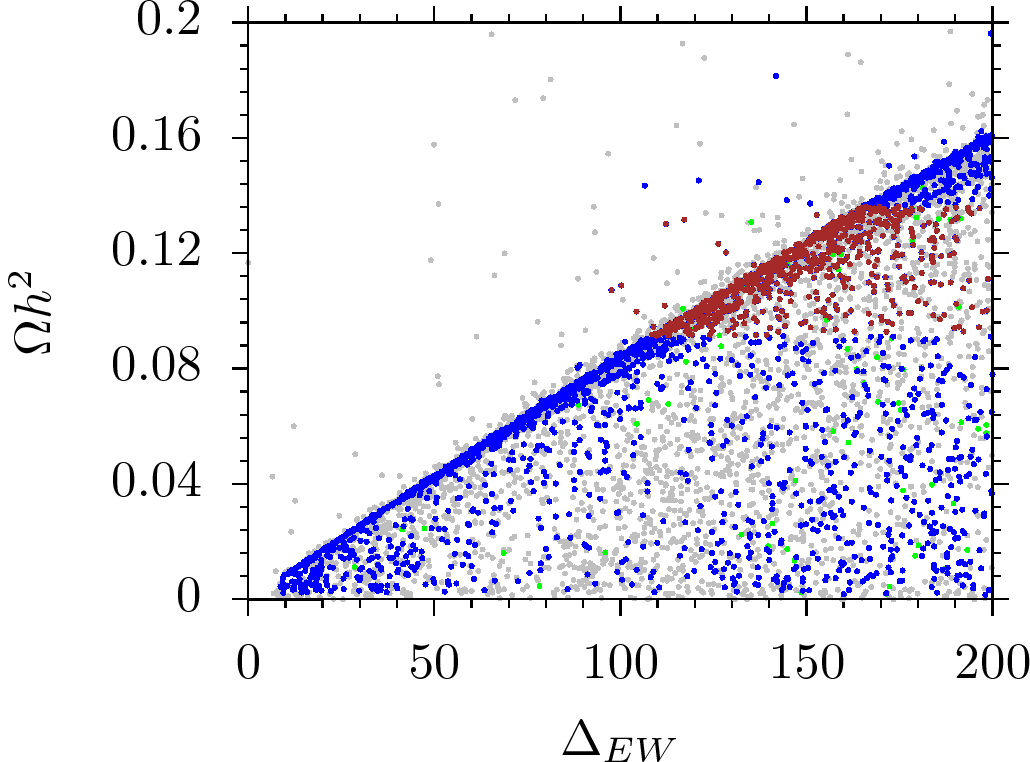}}
\caption{Plots in the $C_{b\tau}-\Delta_{EW}$  and the $\Omega h^2 - \Delta_{EW}$ plane. Color coding is the same as Figure \ref{fig2}. }
\label{fig3}
\end{figure}

We next discuss fine-tuning through the $C_{b\tau}-\Delta_{EW}$ and the $\Omega h^2 - \Delta_{EW}$ plots in Figure \ref{fig3}. The color coding is the same as Figure \ref{fig2}. If dark matter is to be solely composed of Higgsinos, then the WMAP bound imposes a lower bound of $\sim$ 100 on $\Delta_{EW}$, as seen from the $C_{b\tau}-\Delta_{EW}$ plot. However, if we allow for multi-component dark matter, then solutions with $\Delta_{EW}$ as low as about 10 can also be found. 

\section{Sparticle Mass Spectrum and Fine-Tuning}
\label{sec:spec}

\begin{figure}[h!]
\centering
\subfigure{\includegraphics[scale=1]{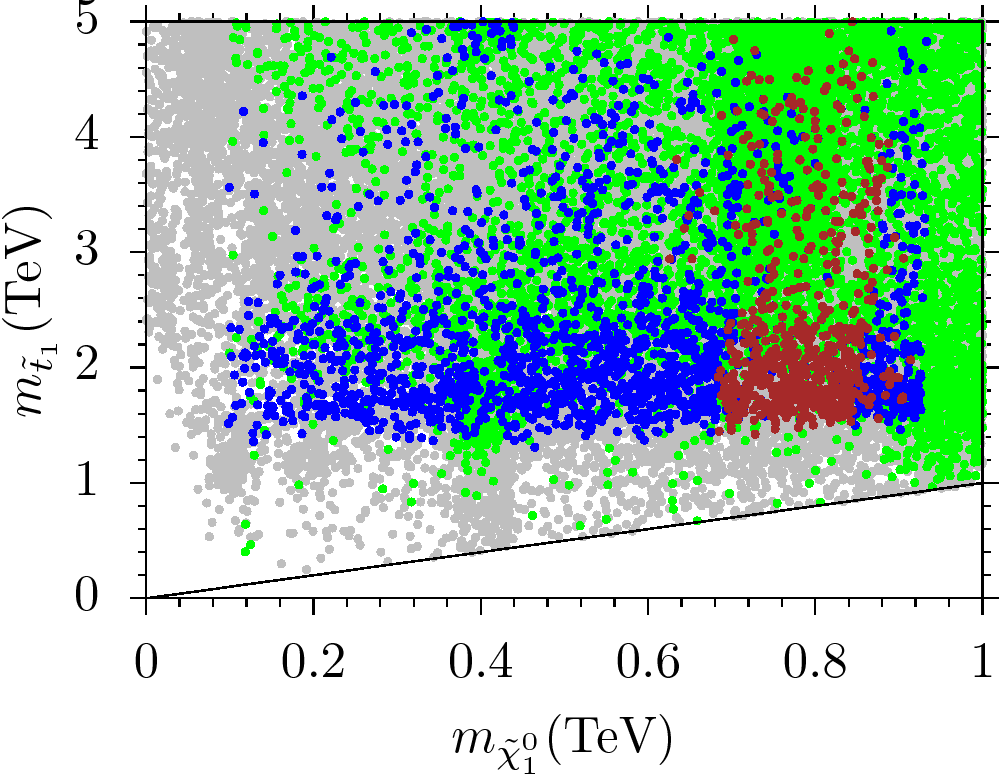}}
\subfigure{\includegraphics[scale=1]{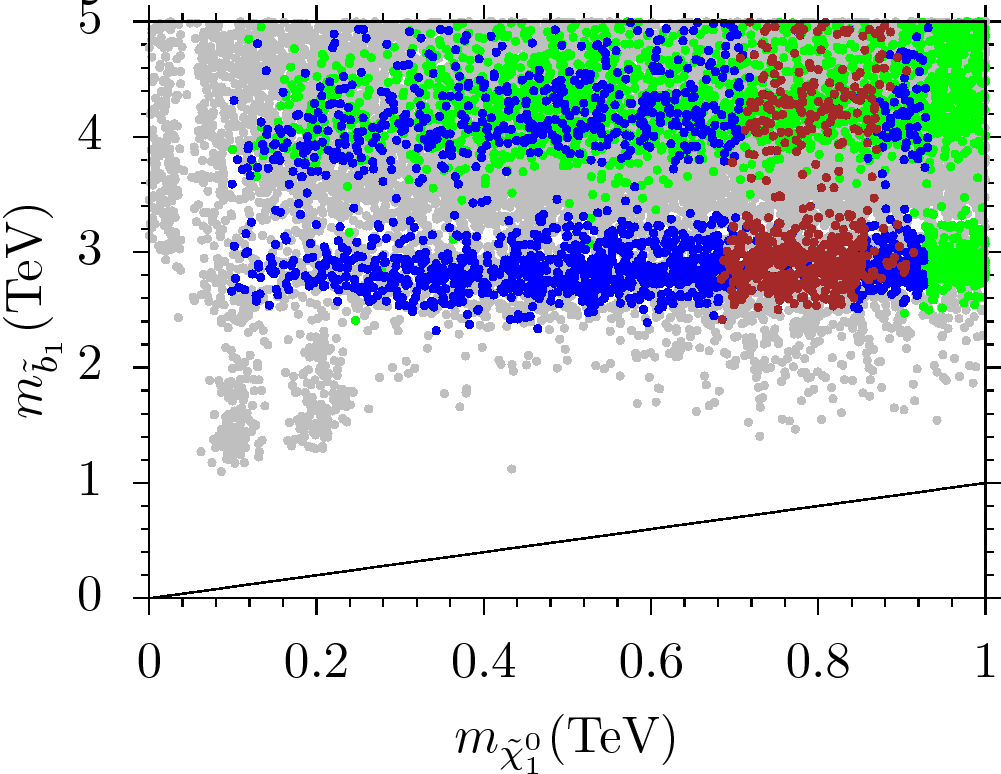}}
\subfigure{\includegraphics[scale=1]{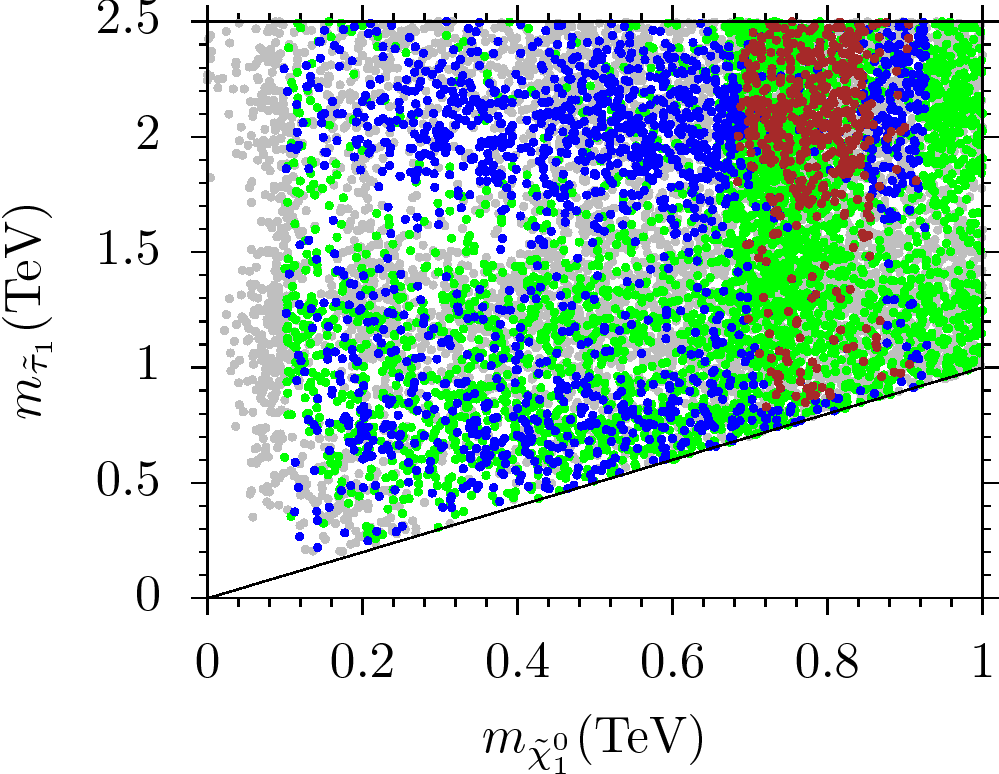}}
\subfigure{\includegraphics[scale=1]{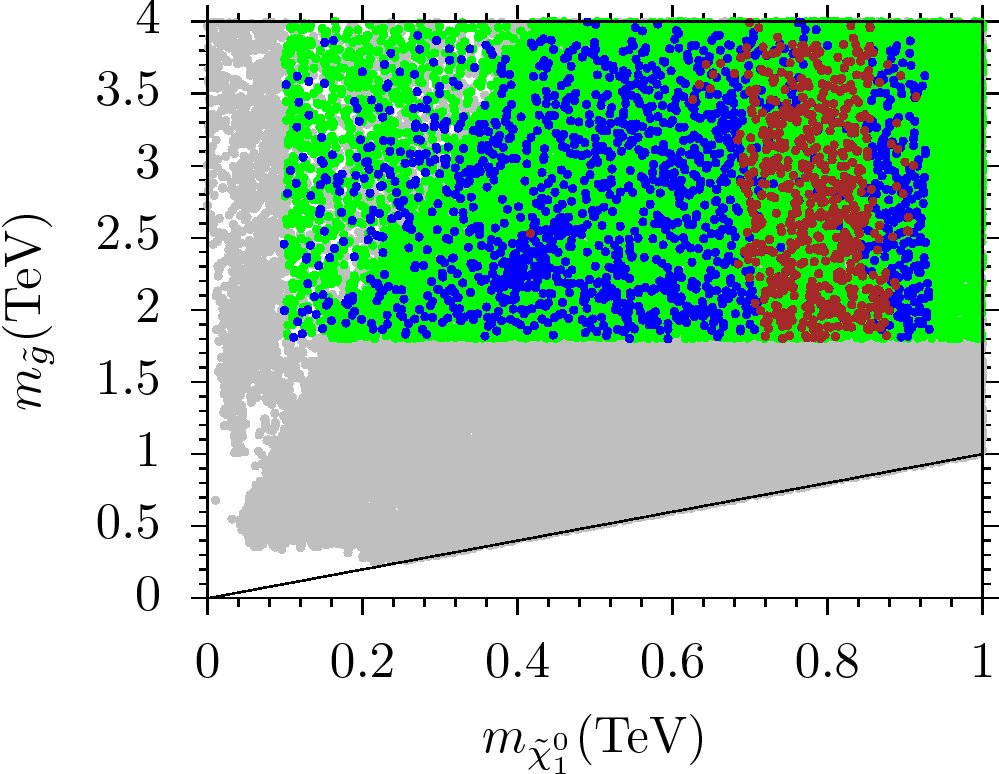}}
%\end{figure}
%\begin{figure}
\subfigure{\includegraphics[scale=1]{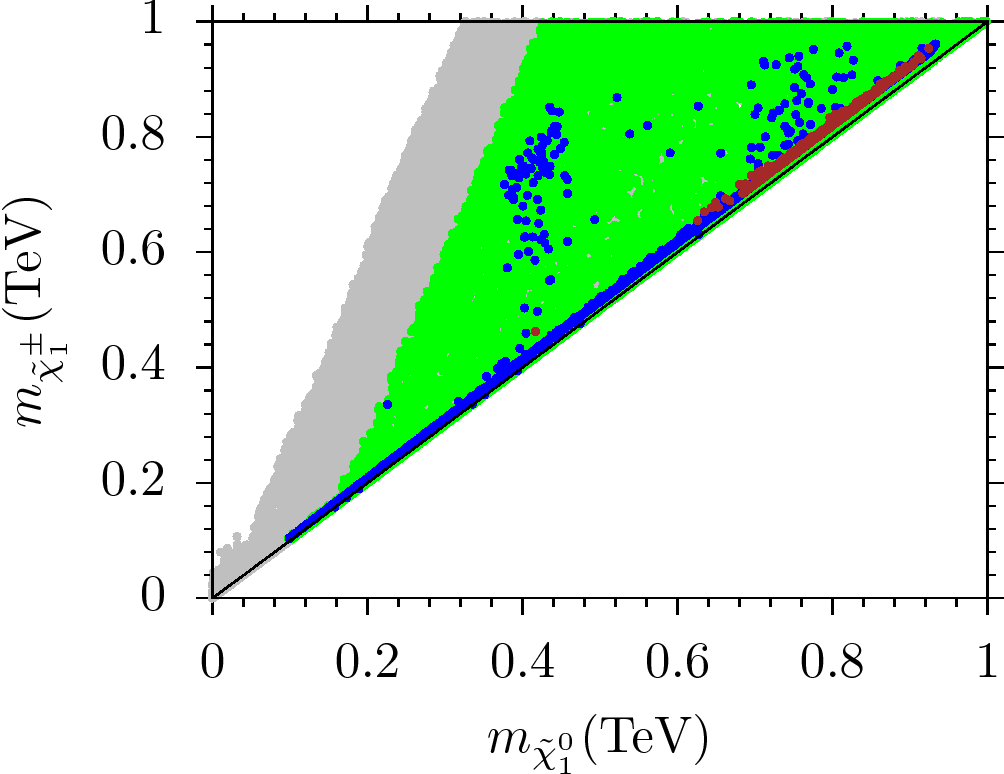}}
%\subfigure{\includegraphics[scale=1]{QYU_422_pos_mA_neut.png}}
\caption{Plots in the $m_{\tilde{t}_{1}}-m_{\tilde{\chi}_{1}^{0}}$, $m_{\tilde{g}}-m_{\tilde{\chi}_{1}^{0}}$, $m_{\tilde{\tau}_{1}}-m_{\tilde{\chi}_{1}^{0}}$ and $m_{\tilde{\chi}_{1}^{\pm}}-m_{\tilde{\chi}_{1}^{0}}$. Color coding is the same as in Figure \ref{fig2}. The lines depict the regions where the sparticle and the LSP are nearly mass degenerate.}
\label{fig4}
\end{figure}

In this section, we discuss the mass spectrum compatible with the $b-\tau$ QYU. Figure \ref{fig4} displays  plots in the $m_{\tilde{t}_{1}}-m_{\tilde{\chi}_{1}^{0}}$, $m_{\tilde{g}}-m_{\tilde{\chi}_{1}^{0}}$, $m_{\tilde{\tau}_{1}}-m_{\tilde{\chi}_{1}^{0}}$ and $m_{A}-m_{\tilde{\chi}_{1}^{0}}$ planes. The color coding is the same as in Figure \ref{fig2}. The lines depict the regions where the sparticle and the LSP are nearly degenerate in mass. We see that the mass of the stop squarks is $\gtrsim 1.5$ TeV, while the gluino mass is bounded below by the current LHC lower bound of $1.8$ TeV. The sbottom is heavier with mass $\gtrsim$ 2.5 TeV. We also show a plot of $m_{\tilde{\tau_1}}$ vs $m_{\tilde{\chi}_1^0}$, from which we note that there exist solutions with NNLSP stau of mass as low as 200-250 GeV.

\begin{figure}[ht!]
\centering
\subfigure{\includegraphics[scale=1]{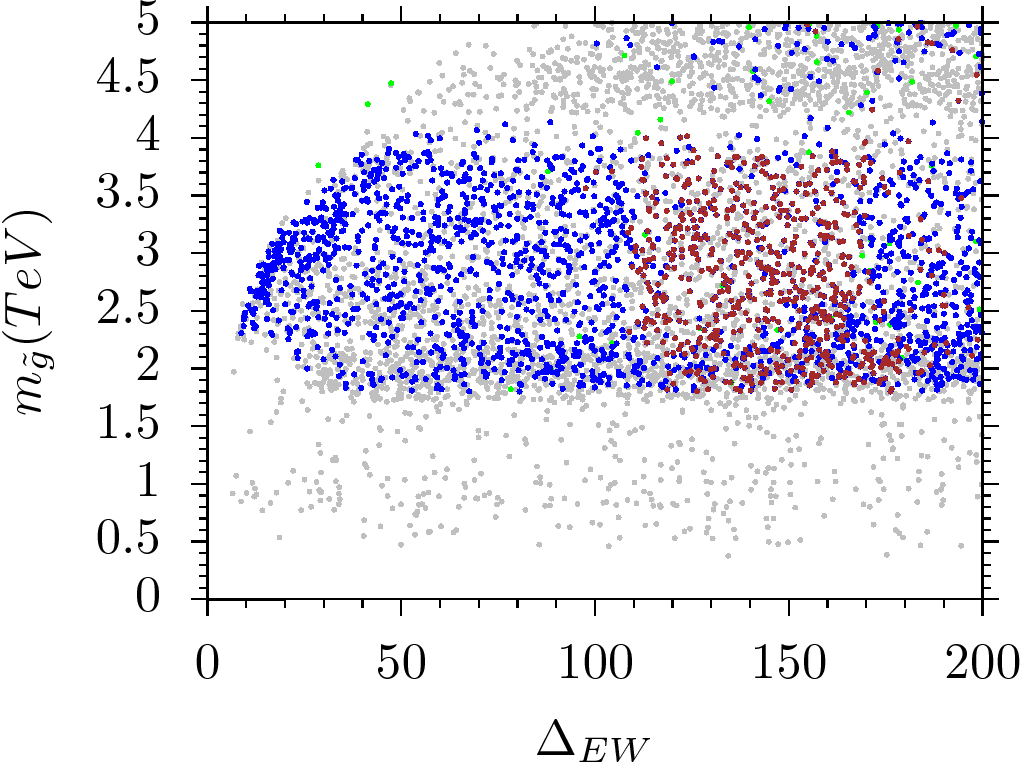}}
\subfigure{\includegraphics[scale=1]{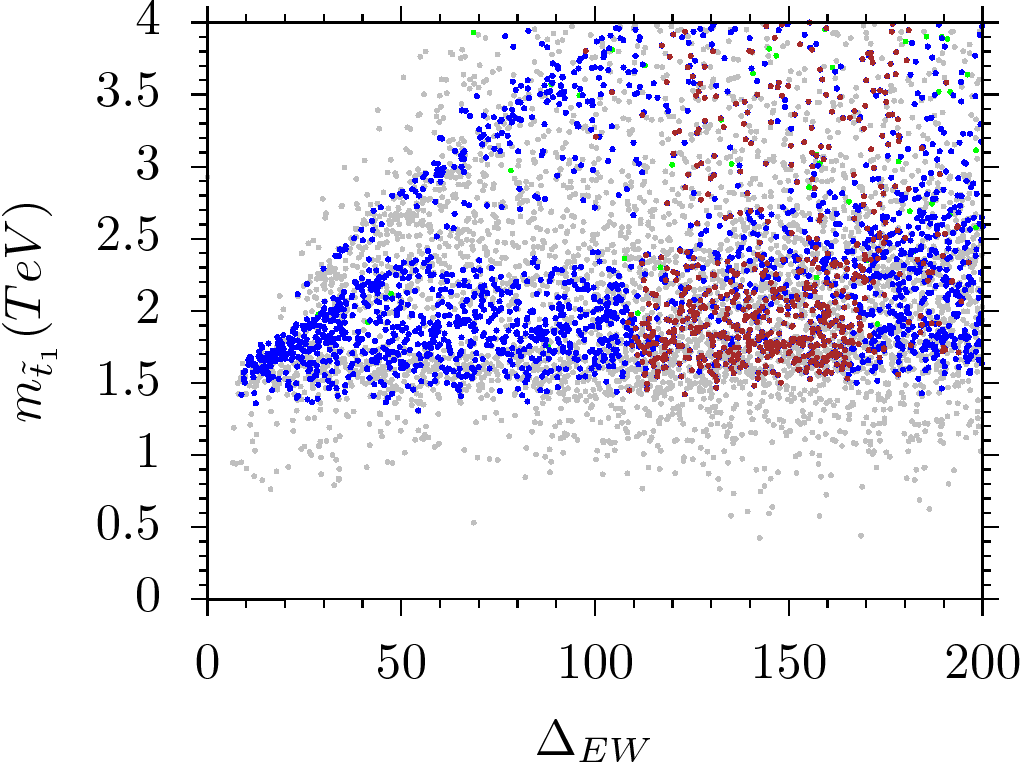}}
\caption{Plots in the $m_{\tilde{g}}-\Delta_{EW}$ and $m_{\tilde{t}_1}-\Delta_{EW}$  planes. The color coding is the same as Figure \ref{fig2}. The LHC lower bound on the gluino mass $m_{\tilde{g}} \geq 1.8$ TeV has been imposed.}
\label{fig5}
\end{figure}

We display plots between the lightest two colored supersymmetric particles, the stop and the gluino, and the fine-tuning parameter $\Delta_{EW}$. The color coding is the same as in Figure \ref{fig2}. We note that it is possible to have a gluino with mass $\sim$ 2-4 TeV for $\Delta_{EW}$ as low as 30 or so \cite{Baer:2016wkz}. We further observe that the stop mass can be as low as 1.5 TeV for the entire range of $\Delta_{EW}$ that we consider. In order to be consistent with the measured mass of the Higgs boson at the LHC, we require either a heavy stop, or large SSB trilinear scalar coupling or a suitable combination of the two. Large SSB trilinear scalar coupling, however, leads to higher fine-tuning \cite{Cici:2016oqr}. 
\section{Higgsino Dark Matter and Direct Detection}
\label{sec:DM}

\begin{figure}[ht!]
\centering
\subfigure{\includegraphics[scale=1]{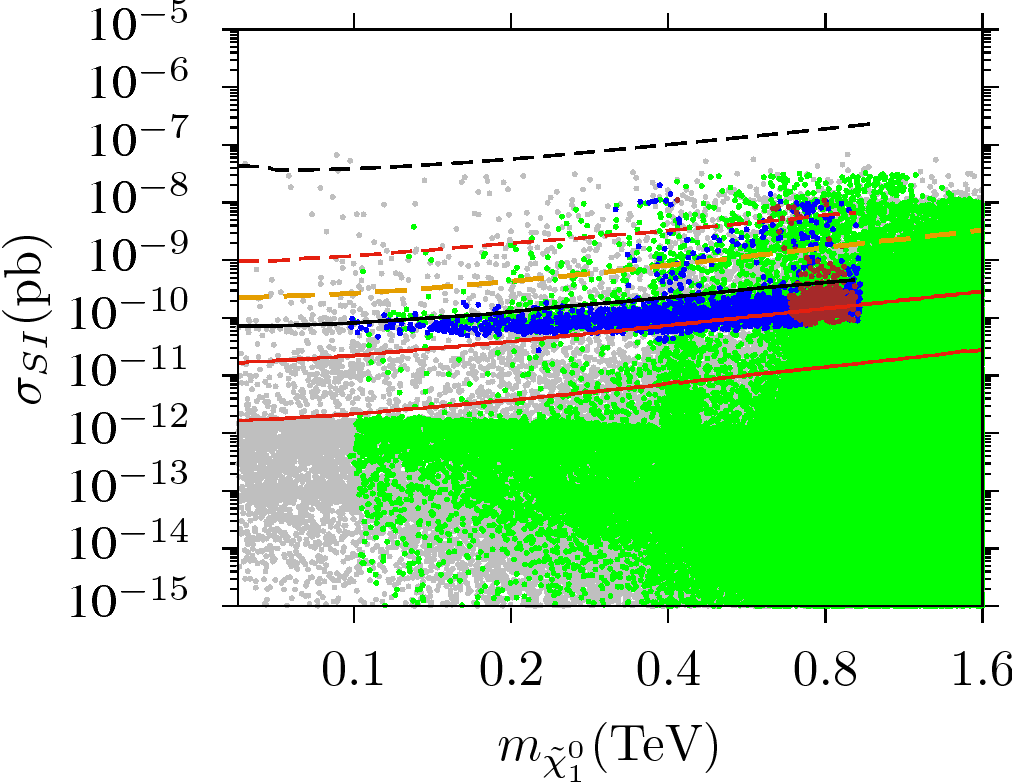}}
\subfigure{\includegraphics[scale=1]{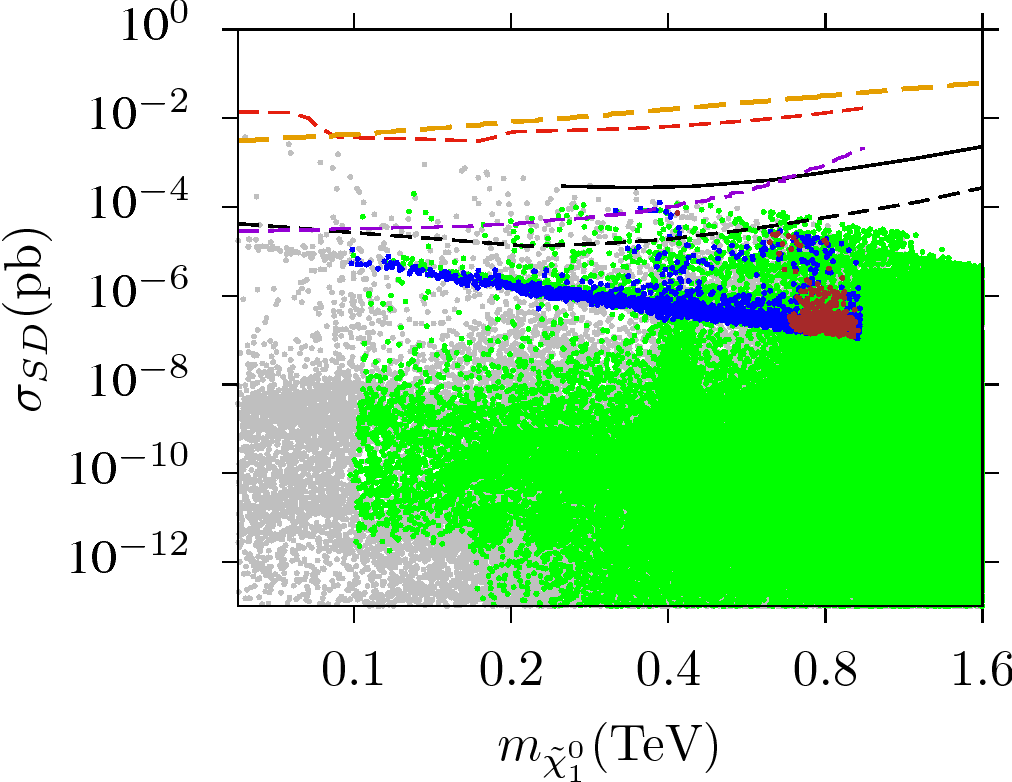}}
\caption{Plots in the $\sigma_{SI} - m_{\tilde{\chi}_{1}^{0}}$ and $\sigma_{SD} - m_{\tilde{\chi}_{1}^{0}}$ planes. Color coding is the same as in Figure \ref{fig1}. In the $\sigma_{SI} - m_{\tilde{\chi}_{1}^{0}}$ plane, the dashed (solid) black line represents the current (future) results from the SuperCDMS experiment \cite{Brink:2005ej}, and dashed (solid) red line(s) shows the current (future) results from the Xenon experiment \cite{Aprile:2015uzo}. The brown solid line shows the latest result from the LUX experiment \cite{Akerib:2016vxi}.  In the $\sigma_{SD} - m_{\tilde{\chi}_{1}^{0}}$ plane, the current upper bounds are set by the Super-K \cite{Tanaka:2011uf}, indicated by the red dashed line, and the IceCube DeepCore by the black dashed (solid) line for its current and expected future results. In addition, the purple dashed line is the limit set by the CMS analysis \cite{Chatrchyan:2012me}, while the brown dashed line represents the latest results from the LUX experiment\cite{Akerib:2016lao}. }
\label{fig7}
\end{figure}

This section discusses the DM implications of $b-\tau$ QYU in the light of current and expected future results from the direct detection experiments. Figure \ref{fig7} shows the results with plots in the $\sigma_{SI} - m_{\tilde{\chi}_{1}^{0}}$ and $\sigma_{SD} - m_{\tilde{\chi}_{1}^{0}}$ planes. The color coding is the same as in Figure \ref{fig2}. In the $\sigma_{SI} - m_{\tilde{\chi}_{1}^{0}}$ plane, the dashed (solid) black line represents the current (future) results from the SuperCDMS experiment \cite{Brink:2005ej}, and dashed (solid) red line(s) shows the current (future) results from the Xenon experiment \cite{Aprile:2015uzo}. The brown solid line is the latest result from the LUX experiment \cite{Akerib:2016vxi}. In the $\sigma_{SD} - m_{\tilde{\chi}_{1}^{0}}$ plane, the current upper bounds are set by Super-K \cite{Tanaka:2011uf} (red dashed line) and the IceCube DeepCore indicated by black dashed (solid) line for its current(future) results. In addition, the purple dashed line is the limit set by the CMS analyses \cite{Chatrchyan:2012me}, while the brown dashed line represents the latest result from the LUX experiment\cite{Akerib:2016lao}.

As can be seen from the $\sigma_{SI} - m_{\tilde{\chi}_{1}^{0}}$ plane, the DM scattering rate on nuclei yields relatively large cross-sections ($\sim 10^{-8}$ pb). These solutions involve higgsino-like DM, and the large cross-section comes from the Yukawa interactions between the higgsinos and quarks in nuclei.  Although some of these solutions are excluded by the current results from the LUX experiment, they will be further tested by the SuperCDMS experiment. The  penultimate solid red line represents the future results from Xenon 1T, which according to present plans, will be reached in 2017. The last solid red line is the projected result from the Xenon experiment over the next 20 years. The spin-dependent scattering results are shown in the $\sigma_{SD} - m_{\tilde{\chi}_{1}^{0}}$ plane, and all solutions are allowed by the current and future reaches of the experiments. 

Finally, we present a table of six benchmark points, which exemplify our findings. The points chosen are consistent with the experimental constraints in \ref{sec:scan}. The lowest value of $\Delta_{EW}$ that we found was 9.6 , with a LSP mass of 207 GeV, displayed in Point 1. Note that since the relic LSP density is about 10 \% of the desired DM abundance, we posit that an additional DM component such as axion is also present. Points 2, 3 and 4 have progressively heavier LSPs which form a larger component of DM, but they require higher fine-tuning as measured by $\Delta_{EW}$. Point 4 with an LSP mass of 688 GeV is the lightest higgsino DM compatible with the WMAP bound and we do not need any other dark matter component in this case.  This point also corresponds to the lowest value of $\Delta_{EW} \approx 110$ compatible with the WMAP bound for higgsino DM. Point 5 displays a pure higgsino DM solution with the central value of relic density ($\Omega h^2$ = 0.113). 

\begin{table}[h!] \hspace{0.0cm}
\centering
\scalebox{0.7}{
\begin{tabular}{|c|ccccc|}
\hline
\hline
&&&&&\\
                & Point 1 & Point 2 & Point 3 & Point 4 & Point 5  \\ 
              &&&&&\\
\hline
$m_{0}$       & 3074 & 2375 & 2057 & 2745 & 2133  \\
$M_{1} $      & 3941 & 3607 & 3350  & 3557 & 3387   \\
$M_{2} $      & 5845 & 5009 & 4536  & 5069 & 4545    \\
$M_{3} $      & 1084 & 1504 & 1571  & 1289 & 1654 \\
$M_{H_d}$     & 615  & 725   & 800 & 609  & 752  \\
$M_{H_u}$     & 1759 & 2025  & 2054  & 2152 & 2121  \\
$\tan\beta$   & 44.7 & 42.9  & 40.5  & 44.3 & 42.3     \\
$A_0/m_{0}$   & -0.57 & -0.62 & -0.66 & -0.72 & -0.65 \\
$\mu$         & \textbf{212.3} & \textbf{333.9}  & \textbf{472}  & \textbf{699}  & \textbf{786}   \\
$\Delta EW$   & \textbf{9.8} & \textbf{27.4}  & \textbf{48.8}  & \textbf{109.6}  & \textbf{137.5}   \\
\hline
$m_h$         & 123.2 & 123.5  & 123.3 & 123.6 & 123.4  \\
$m_H$         & 1895  & 1586  & 1639 & 1244  & 1447   \\
$m_A$         & 1883  & 1576   & 1629  & 1236  & 1437   \\
$m_{H^{\pm}}$ & 1897  & 1588   & 1642  & 1248  & 1450   \\

\hline
$m_{\tilde{\chi}^0_{1,2}}$
              & \textbf{207.3}, \textbf{209.4} & \textbf{326.8}, \textbf{329.3}    & \textbf{460.7}, \textbf{463.6}  & \textbf{688.1}, \textbf{690.8} & \textbf{769.7}, \textbf{772.9}    \\

$m_{\tilde{\chi}^0_{3,4}}$
              & 1792, 4856 & 1629, 4138     & 1509, 3739 & 1614, 4206 & 1529, 3749   \\

$m_{\tilde{\chi}^{\pm}_{1,2}}$
              & \textbf{215.7}, 4845 & \textbf{338.2}, 4120  & \textbf{475.1}, 3720  &   \textbf{706.2}, 4189 & \textbf{789.5}, 3728  \\

$m_{\tilde{g}}$  & \textbf{2487} & \textbf{3291} & \textbf{3407}   & \textbf{2884} & \textbf{3571}   \\
\hline $m_{ \tilde{u}_{L,R}}$
                 & 5099, 3754 & 4719, 3713  & 4460, 3604  & 4740, 3706 & 4585, 3762     \\
$m_{\tilde{t}_{1,2}}$
                 & 1584, 4189 & 1939, 3910  & 2002, 3728   & 1760, 3845 & 2171, 3823  \\
\hline $m_{ \tilde{d}_{L,R}}$
                 & 5100, 3646 & 4720, 3616 & 4460, 3513  & 4740, 3606  & 4585, 3672  \\
$m_{\tilde{b}_{1,2}}$
                 & 2788, 4231 & 2901, 3936 & 2923, 3748  & 2795, 3871 & 3031, 3838  \\
\hline
$m_{\tilde{\nu}_{e,\mu}}$
                 & 4843 & 4005 & 3579 & 4262 & 3627    \\
$m_{\tilde{\nu}_{\tau}}$
                 & 4582 & 3798  & 3407 & 4008 & 3432 \\
\hline
$m_{ \tilde{e}_{L,R}}$
                & 4838, 3373 & 4002, 2684 & 3578, 2357   & 4258, 3002 & 3626, 2427  \\
$m_{\tilde{\tau}_{1,2}}$
                & 2447, 4560 & 1941, 3784 & 1744, 3397   & 2131, 3992 & 1742, 3423 \\
\hline

$\sigma_{SI}({\rm pb})$
                & $0.66 \times 10^{-10}$ & $0.10 \times 10^{-09} $ & $0.14\times 10^{-09}$  & { $0.17\times 10^{-09}$} & $ 0.23 \times 10^{-09} $ \\

$\sigma_{SD}({\rm pb})$
                & $0.12 \times 10^{-05} $ & $0.78 \times 10^{-06} $   & $0.58\times 10^{-06}$   & $0.28\times 10^{-06}$ & $ 0.32 \times 10^{-06}$ \\

$\Omega h^{2}$  & 0.008 & 0.02  & 0.041   & 0.0914 & 0.113 \\
\hline
&&&&&\\
$y_{t,b,\tau}(M_{\rm GUT}) $ & 0.55, 0.30, 0.41 & 0.54, 0.30, 0.39  & 0.54, 0.28, 0.36   & 0.55, 0.32, 0.42  & 0.54, 0.30, 0.39  \\
&&&&&\\
$ C $  & 0.08 & 0.07  & 0.07 & 0.08 & 0.08 \\
\hline
\hline
\end{tabular}}
\caption{Benchmark points consistent with the experimental constraints mentioned in Section \ref{sec:scan}. All masses are given in GeV. All points involve essentially 100\% higgsino dark matter.}
\label{ISAJET} 
\end{table}

\newpage
\section{Conclusion}
\label{sec:conc}

We have explored how light ($\lesssim$ 1 TeV) higgsinos can arise in supersymmetric SU(4)$_C \times$ SU(2)$_L \times$ SU(2)$_R$ models which exhibit quasi b-$\tau$ Yukawa unification. We also require that the electroweak fine tuning measure $\Delta_{EW}\lesssim$ 200. In the colored sector the stop mass is greater than 1.5 TeV or so. The first two family squarks are considerably heavier approaching 10 TeV in some cases. The gluino mass is estimated to lie in the 2-4 TeV range, which poses the important question: Will the LHC find the gluino?

\newpage
%\vspace{0.3cm}
\noindent {\bf Acknowledgments}
This work is supported in part by the DOE Grant DE-SC0013880 (A.H. and Q.S.), and the Scientific and Technological Research Council of Turkey (TUBITAK) Grant no. MFAG-114F461 (C.S.\"{U}.). This work used the Extreme Science and Engineering Discovery Environment (XSEDE), which is supported by the National Science Foundation grant number OCI-1053575. Part of the numerical calculations reported in this paper were performed at the National Academic Network and Information Center (ULAKBIM) of TUBITAK, High Performance and Grid Computing Center (TRUBA Resources).

%\newpage

%\include{bibliography}

\end{document}